\documentclass[a4j,10pt,onecolumn,oneside,notitlepage,final]{article}
\usepackage[dvips]{graphicx}

\def\br{\boldmath{r}}
\def\bR{\boldmath{R}}
\def\pat{\partial}
\def\half{\frac{1}{2}}

\begin{document}

\title{Long-range attractive tail of colloid interaction}
\author{Ikuo S. Sogami\footnote{sogami@cc.kyoto-su.ac.jp}
\smallskip
\\
{\it Maskawa Institute, Kyoto Sangyo University, Kyoto 603-8555}}

\date{}

\maketitle

\begin{abstract}
To describe rich phenomena of dilute dispersions of highly-charged colloid particles,
it is necessary to modify the long-range electric part of the DLVO potential. The
screened Coulomb potential of pure repulsion obtained from the Helmholtz
free energy in the DLVO theory is replaced by a new screened electric potential with
long-range attractive tail derived from the generalized Gibbs free energy which is
identified with a total sum of chemical potentials.
\end{abstract}

\bigskip

Colloidal dispersions display a variety of phenomena~\cite{IseSogami} such as ordering formation
and phase transitions. Diversity of those phenomena has its origin in a subtle characteristic of
interaction potential of colloid particles, which changes its medium- and long-range properties
depending on salts and particle concentrations. The DLVO potential~\cite{DL,VO}
consisting of the van der Waals attraction and the screened Coulomb repulsion which is derived
from the Helmholz free energy $F$ of the system can explain the Schulze-Hardy empirical rule
for coagulation of colloid particles. This success of short-range characteristics of
the colloid interaction had granted it the position of the standard theory of colloid
dispersions. However, it turns out impossible to describe rich phenomena of medium- and
long-range nature of colloid dispersions in the DLVO theory.  In this note, we show that
a generalized concept of the Gibbs free energy $G$ is more appropriate than the Helmholtz
free energy $F$ to describe the chemical as well as thermal equilibrium of the colloidal
dispersion and prove that a pair potential derived from $G$ possesses a favorable property
which can explain characteristics of the dispersions over all ranges. 

To maintain analytical derivations of thermodynamic functions of colloid systems, we accept
the mean field theory developed originally by Debye and H\"{u}ckel~\cite{DH,FG,McQ}
(DH) to investigate the strong electrolyte. In their theory, the electric part of
the internal energy $E^{el}$ is calculated from the mean electric potential $\psi(\br)$
which is determined by solving the linearized Poisson-Boltzmann (PB) equation.
Thanks to homogeneousness of the system, the thermodynamic relations
\begin{equation}
      E^{el} = F^{el} - T\frac{\pat F^{el}}{\pat T}
      \label{LT:EF}                                                                                  
\end{equation}
and
\begin{equation}
      G^{el} = F^{el} - V\frac{\pat F^{el}}{\pat V}
      \label{LT:GF}                                                                                   
\end{equation}
can be applied step by step to derive the electric parts of the Helmholtz free energy
$F^{el}$ and then the Gibbs free energy $G^{el}$.

It is also possible to calculate $G^{el}$ by summing up the chemical potentials of all ions as
\begin{equation}
        G^{el} = D F^{el} = \sum_j N_j\frac{\pat F^{el}}{\pat N_j}
        \label{SumChemSE}                                                                                   
\end{equation}
in which $N_j$ is the number of ions of j-th species with valence $z_j$. Here, we introduce
the differential operator
\begin{equation}
        D = \sum_j N_j\frac{\pat}{\pat N_j} 
        \label{DerivativeSE}                                                                                  
\end{equation}
for convenience in later arguments on the colloid interaction. Note that the operator
$D$ is compatible with the condition of electric charge neutrality
\begin{equation}
        \sum_j z_j N_j = 0
        \label{NeutralitySE}                                                                                   
\end{equation}
in the sense that this relation is invariant under its action.
 
The DH method has previously been adopted in theories of colloid dispersions. First,
Levine and Dube~\cite{LD} derived a pair potential of particles with a medium-range
repulsion and long-range attraction from the internal energy $E^{el}$. Subsequently,
Derjaguin-Landau~\cite{DL} and Verway-Overbeek~\cite{VO} utilized Eq.\,(\ref{LT:EF}) to derive
$F^{el}$ from $E^{el}$ and obtained a purely repulsive pair potential. Note that the Legendre
transformation in Eq.\,(\ref{LT:EF}) is applicable because colloid dispersions which are
in a thermal bath of solvent can be regarded as homogeneous with respect to the temperature $T$.
The final step from $F^{el}$ to $G^{el}$ is still very problematic for a colloid.
The equivalence of two routes to the Gibbs free energy by means of the Legendre
transformation in Eq.\,(\ref{LT:GF}) (route L) and the sum of the chemical potentials
in Eq.\,(\ref{SumChemSE}) (route C), which holds in the DH theory of strong electrolytes,
seems not to hold~\cite{Sogami,SogamiIse,SSS} for the inhomogeneous colloid dispersions.
We will carry out below explicit derivations along these two routes and
prove that, while the route C gives us a favorable pair potential,  the route L leads
us to an unphysical result.  

To examine the electric interaction, let us focus our attention on a dilute dispersion
of colloid particles with large analytic surface charges. The large analytic charges
act to attract and to condense counterions near the surfaces and the condensed counterions
work to screen the analytic charges. Hereafter, we call such screened charge the
{\it effective charge} and the colloid particle with the effective charge simply
the {\it particle}. In the dilute dispersion where particles have static configurations,
the gasses of small ions are postulated to be in the thermal and chemical equilibrium
described by the mean electric potential $\psi(\br)$ satisfying the linearized PB equation. 

For simplicity, we consider a monodisperse dispersion consisting of particles with
effective valence $Z$ and radius $a$. The internal energy $E^{el}$ is given by
integrating the electric energy density over the dispersion as follows:
\begin{equation}
         E^{el} = \half \sum_j z_je \int \psi({\br}) n_j({\br}) d{\br}
                + \half \sum_n Ze \int \psi({\br}) \rho({\br}) d{\br} 
         \label{Eintegral}                                                                               
\end{equation}
where $n_j({\br})$ is the Boltzmann distribution of the small ions and $\rho({\br})$ is
the charge distribution function of the particle at $\bR_n$. Note that the mean electric
potential is defined except for arbitrary constants.  Namely,  the theory is invariant
under the shift $\psi(\br) \rightarrow \psi(\br) + c$ with an arbitrary constant $c$.
By applying this replacement to Eq.\,(\ref{Eintegral}), we obtain the constraint of charge neutrality 
\begin{equation}
        \sum_j z_jN_j + NZ = 0
        \label{NeutralityCD}                                                                                
\end{equation}
where $N$ is the number of particles. Note that, while $N$ is fixed, $Z$ is assumed
to be variable in the present postulation.

Substituting the mean electric potential determined by solving the linearized PB equation
into Eq.\,(\ref{Eintegral}) and integrating out the degrees of freedom of the electric field
and small ions, we find the following expression
\begin{equation}
        E^{el} = E_0 + \half\sum_{m\neq n}U^{E}(R_{mn}) + \sum_n V_n^{E}
        \label{Eexpand}                                                                                
\end{equation}
in which $R_{mn}=|\bR_m - \bR_n|$ is the center to center distance between
the m-th and n-th particles.  The pair potential takes the form~\cite{IseSogami}
\begin{equation}
        U^{E} = \frac{Z^{\ast 2}e^2}{\epsilon}\left[\,\frac{\kappa a\coth(\kappa a)}{R}
                -\half \kappa\,\right] e^{-\kappa R} 
        \label{Epot}                                                                              
\end{equation}
where
\begin{equation}
        Z^{\ast} = Z\frac{\sinh(\kappa a)}{\kappa a}
        \label{Zstar}                                                                                
\end{equation}
and $\kappa$ is Debye's screening parameter defined by
\begin{equation}
        \kappa^2 = \frac{4\pi e^2}{\epsilon k_B TV}\sum_jz_j^2N_j .
        \label{Debye}                                                                                
\end{equation}
The quantity $V^E_n$ in Eq.\,(\ref{Eexpand}) is the self-energy of the n-th particle in
the ionic atmosphere of the dispersion.

 In parallel with the relation in Eq.\,(\ref{Eexpand}),  it is possible to express the free
energy $F^{el}$ as
\begin{equation}
       F^{el} = F_0 + \half\sum_{m\neq n}U^{F}(R_{mn}) + \sum_n V_n^{F} 
       \label{Fexpand}                                                                               
\end{equation}
in terms of the Helmholtz pair potential and self-energy. As already remarked above,
the temperature $T$ is a relevant thermodynamic variable for the colloid dispersion.
Therefore, we can obtain the Helmholtz free energy $F^{el}$ from $E^{el}$ by way of
the relation in Eq.\,(\ref{LT:EF}). Thus, the two kinds of pair potentials satisfy
the differential equation
\begin{equation}
       U^E(R) = U^F(R) - T\frac{\pat U^F(R)}{\pat T} 
       \label{Diff:EF}                                                                           
\end{equation}
which can readily be solved, resulting in
\begin{equation}
        U^{F} = \frac{Z^{\ast 2}e^2}{\epsilon}\frac{1}{R} e^{-\kappa R}
        \label{Fpot}                                                                                
\end{equation}
Here it should be pointed out that the pair potentials in Eqs.\,(\ref{Epot}) and (\ref{Fpot})
are approximately equivalent to the ones derived, respectively, by Levine-Dube and DLVO.

Now, the problem is to find a generalized concept of the {\lq}{\lq}Gibbs free energy{\rq}{\rq}      which is suitable to describe the inhomogeneous colloidal dispersions.  The free energy $G^{el}$
can also be expressed as follows:
\begin{equation}
        G^{el} = G_0 + \half\sum_{m\neq n}U^{G}(R_{mn}) + \sum_n V_n^{G} 
        \label{Gexpand}                                                                             
\end{equation}
where $U^G(R)$ is the Gibbs pair potential to be derived. It must be examined which
one of the routes L and C lead us to a suitable pair potential. 

To follow route L, the Legendre transformation in Eq.\,(\ref{LT:GF}) has to be applied to
the free energies $F^{el}$ and $G^{el}$. Consequently, the differential equation
\begin{equation}
         U^G(R) = U^F(R) - V\frac{\pat U^F(R)}{\pat V}
         \label{Diff:GF}                                                                                
\end{equation}
is derived for the pair potentials $U^F(R)$ and $U^G(R)$. To solve this equation,
we must observe the facts that the equations in Eqs.\,(\ref{Diff:EF}) and (\ref{Diff:GF}) are
identical under the replacement of ( $T$ and $U^F(R)$ ) by ( $V$ and $U^G(R)$ ) and that
the Debye's screening parameter through which the pair potentials depend on the variables
$T$ and $V$ is symmetric with respect to these variables. Therefore, the Gibbs pair potential
satisfying Eq.\,(\ref{Diff:GF}) is proved to be identical with the electric pair potential $U^E(R)$
given in Eq.\,(\ref{Epot}). Namely, the route L of the Legendre transformation is destined to
result in the unacceptable result $U^G(R)=U^E(R)$. When combined with the van der Waals
potential, this pair potential fails to explain the Schulze-Hardy rule.
 
The source of failure of the route L seems to be traced back to the fact that $V$ in
Eqs.\,(\ref{LT:GF}) and (\ref{Debye}) is not a suitable thermodynamic variable.
In Eqs.\,(\ref{LT:EF}), (\ref{LT:GF}) and (\ref{Debye}), the variables $T$ and $V$
appear exactly symmetric manner. However, while the colloid dispersion is homogeneous
with respect to $T$, it is not with respect to $V$. For the gas of small ions,
the particles form inhomogeneous environment exerting the volume exclusion effect. 
In the theory of colloid dispersions and macroionic systems, the quantity $V$ must be treated,
not a simple variable, but an implicit function of configurations of macroionic particles.
Careless arguments along the route L by treating $V$ as a thermodynamic variable must be
forbidden. It is prudent to renounce Eqs.\,(\ref{LT:GF}) and (\ref{Diff:GF}) including $V$ as
the variable and to interpret $V$ in Eq.\,(\ref{Debye}) simply as a given parameter.

 	To realize what is another thermodynamic variable apart from $T$ in the treatment
along route C, let us apply the neutrality conditions in Eqs.\,(\ref{NeutralitySE}) and
(\ref{NeutralityCD}) to regions with various sizes in arbitrary positions in system.
If the size of the region is not very small, the conditions hold approximately well
for both strong electrolytes and colloid dispersions. So, this argument implies that
the mean electric charges inside sizable regions
can play the role of a thermodynamic variable in ionic suspensions. Therefore,
we are allowed to generalize the scheme along the route C in the DH theory to colloid.  

In case of strong electrolytes, $G^{el}$ is given by the sum of the chemical potentials of
ionic species as in Eq.\,(\ref{SumChemSE}), and the operator $D$ is defined in
Eq.\,(\ref{DerivativeSE}) so as to be compatible with the neutrality condition in
Eq.\,(\ref{NeutralitySE}). Therefore, we should define the differential operator $D$
for colloid dispersions that is compatible with the neutrality condition
in Eq.\,(\ref{NeutralityCD}) by
\begin{equation}
          D = \sum_j N_j\frac{\pat}{\pat N_j} + Z\frac{\pat}{\pat Z} 
          \label{DerivativeCD}                                                                                 
\end{equation}
and the {\lq}{\lq}Gibbs free energy{\rq}{\rq} by
\begin{equation}
          G^{el} = D F^{el} = \sum_j N_j\frac{\pat F^{el}}{\pat N_j} + Z\frac{\pat F^{el}}{\pat Z}
          \label{SumChemCD}
\end{equation}
where the derivative of $F^{el}$ with respect to valence $Z$ should be interpreted as
the chemical potential of the effective valence $Z$ which works to measure the extent
of its fluctuation. 
	In this way, the generalized Gibbs free energy for colloid dispersions is defined,
along route C, as the total sum of the chemical potentials of the small ions and
the effective valence of the particles. Substitution of the expressions in
Eqs.\,(\ref{Fpot}) and (\ref{Gexpand}) into Eq.\,(\ref{SumChemCD}) results readily in
the pair potential
\begin{equation}
     U^{G}(R) = DU^{F}(R)
              = \frac{Z^{\ast 2}e^2}{\epsilon}\left[\,\frac{1+\kappa a\coth(\kappa a)}{R}
                -\half \kappa\,\right] e^{-\kappa R} 
    \label{Gpot}                                                                              
\end{equation}
The chemical and thermal processes advance so as to minimize this pair potential in dilute dispersions of highly charged colloid particles. Three types of pair potentials
$U^{E}(R)$, $U^{F}(R)$ and $U^{G}(R)$ are compared in figure 1.
 
The resultant Gibbs pair potential has a medium-range strong repulsion and a long-range weak
attraction. The repulsive part explains the Schulze-Hardy rule~\cite{IseSogami} for coagulation
and the attractive part describes a variety of phenomena of ordering formation in dilute
dispersions of particles. It is worthwhile recognizing that the chemical potential of the
effective valence plays the role of enhancing the magnitude of the medium-range repulsion.
As shown in figure 2, the pair potential changes its behavior depending sensitively on the concentrations of the particles and added electrolytes through the parameter $\kappa a$.
Evidently, the simple relation
\begin{equation}
        U^{G}(R) = U^{E}(R) + U^{F}(R) 
        \label{GEF}                                                                           
\end{equation}
holds among the three pair potentials. The {\lq}{\lq}Gibbs pair potential{\rq}{\rq} derived
along route C shares the good nature of the pair potentials of the internal and Helmholtz
free energies.

\begin{figure}[h]
 \begin{center}
   \includegraphics[width=55mm]{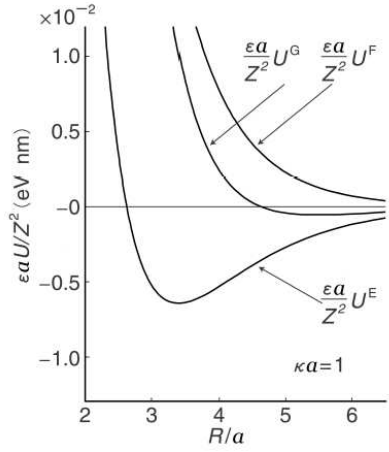}
 \end{center}
 \caption{Comparison of three types of pair potentials.}
\end{figure}

\begin{figure}[h]
 \begin{center}
   \includegraphics[width=65mm]{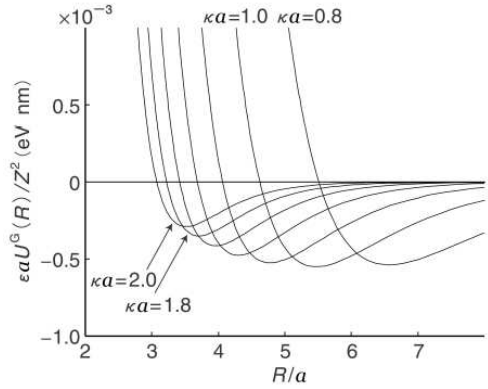}
 \end{center}
 \caption{Behavior of the Gibbs pair potential.}
\end{figure}

In this way, improvement of the long-range features of the DLVO theory has been achieved
without losing the irreplaceable forte in its short-range characteristics. Guided by observations
on the constraint of charge neutrality, we have succeeded in arriving at a generalized concept
of the Gibb free energy for colloid as the total sum of the chemical potentials of the small
ions and the effective valence of the particles. It seems reasonable to evaluate this scheme
as a milestone in trials to improve the fundamental concepts of the thermodynamics of
homogeneous systems so as to be applicable to describe a variety of phenomena in intrinsically inhomogeneous systems. Colloid dispersions which provide us with definite experimental results
will be indispensable and fruitful in studies of  complex and inhomogeneous ionic systems.


\begin{thebibliography}{99}%ŽQl•¶Œ£'̐"'ª1Œ…'È'ç9,3Œ…'È'ç999'É'·'é
\bibitem{IseSogami}
N. Ise and I. S. Sogami, {\it Structure Formation in Solution}, Springer, (2005). 
\bibitem{DL}
B. V. Derjaguin and L. Landau, Acta. Physicochim. URSS {\bf 14} (1941), 633. 
\bibitem{VO}
E. J. W. Verwey and Th. G. Overbeek, Theory of the Stability of Lyophobic Colloids, Elsevier,(1948).
\bibitem{DH}
P. Debye and E. H\"{u}ckel, Physik. Z. {\bf 24} (1923), 185.
\bibitem{FG}
R. H. Fowler and E. A. Guggenheim, Statistical Thermodynamics, CUP, (1935).
\bibitem{McQ}
D. A. McQuarrie, Statistical Mechanics, Harper Collins Publishers, (1976). 
\bibitem{LD}
S. Levine, J. Chem. Phys., 1938, 7, 831, S. Levine and G. P. Dube, Trans. Faraday Soc. 
{\bf 35} (1939), 1125.
\bibitem{Sogami}
I. Sogami, Phys. Lett. A{\bf 96} (1983), 199. 
\bibitem{SogamiIse}
I. Sogami and N. Ise, J. Chem. Phys.{\bf 81} (1984), 6320.  
\bibitem{SSS}
I. S. Sogami, M. Smalley and T. Shinohara, Prog. Theor. Phys. {\bf 113} (2005), 235.
\end{thebibliography}
\end{document}